# Secured Distributed Cognitive MAC and Complexity Reduction in Channel Estimation for the Cross Layer based Cognitive Radio Networks

Niraj Shakhakarmi

Department of Electrical & Computer Engineering, Prairie View A&M University (Texas A&M University System)
Prairie View, Houston, Texas, 77446, USA

**Abstract**

Secured opportunistic Medium Access Control (MAC) and complexity reduction in channel estimation are proposed in the Cross layer design Cognitive Radio Networks deploying the secured dynamic channel allocation from the endorsed channel reservation. Channel Endorsement and Transmission policy is deployed to optimize the free channel selection as well as channel utilization to cognitive radio users. This strategy provide the secured and reliable link to secondary users as well as the collision free link to primary users between the physical and MAC layers which yields the better network performance. On the other hand, Complexity Reduction in Minimum Mean Square Errror (CR-MMSE) and Maximum Likelihood (CR-ML) algorithm on Decision Directed Channel Estimation (DDCE) is deployed significantly to achieve computational complexity as Least Square (LS) method. Rigorously, CR-MMSE in sample spaced channel impulse response (SS-CIR) is implemented by allowing the computationally inspired matrix inversion. Regarding CR-ML, Pilot Symbol Assisted Modulation (PSAM) with DDCE is implemented such the pilot symbol sequence provides the significant performance gain in frequency correlation using the finite delay spread. It is found that CR-MMSE demonstrates outstanding Symbol Error Rate (SER) performance over MMSE and LS, and CR-ML over MMSE and ML.

## 1. Introduction

The concept of cross-layer design is to optimize the overall performance (the routing efficiency, throughput, fairness and delay variance) of Cognitive radio networks with the exchange of information between the different layers of among different applications. Physical layer information exchange with the MAC and network layer has also exhibited superior network performance. The CRN cross layer design involves the MAC and PHY layers, and the central decisions are taken by the CR Manager (CRM) which also has the capability of interaction between different modules. The other two modules in the MAC layer are responsible for efficient dynamic channel allocation among mobile nodes with limiting power constraints. The Channel Sense and Allocation scans spectrum in periodic intervals for possible free unoccupied channels. The next module is called the Power-aware Scheduling aims at a multi-objective joint power control and link scheduling of data frames. Additionally, it also performs the hybrid queuing strategy to achieve fairness among requesting applications. The noise and interference measurement, coded-OFDM transceiver and receiver modules and channel estimation are associated with the PHY layer of each node in the network. Noise and Interference measures the interfering power sensed in each sub-channel due to users in adjacent sub-bands. The coded OFDM transceiver and receiver deals with the PHY layer transceiver and receiver, it can use CR technology to select optimal channel for maximum signal power and hence maximum Signal-to-Noise Ratio (SNR) at minimal Bit Error Rate (BER).

The channel estimation approximates the fading condition of the channel as well as the channel error rates. The CE shares the cross-layer information with the CRM to select the best link for data transmission among the adjacent one hop neighbors. The Decision Directed Channel Estimation (DDCE) scheme is preferred for CR in both information symbols and the pilot symbols for channel estimation. Least Square method has low complexity but the energy-fluctuations associated with the near-Gaussian distributed subcarrier-related samples render the poor performance for decision directed channel in OFDM. On the other hand, MMSE in DDCE provide significant performance over LS but complexity need to be mitigated the intensive matrix inversion operation. Similarly, the Maximum Likelihood (ML) DDCE for CR needs novel time-frequency interpretation for complexity reduction.

The spectrum mobility is the ability of a distributed cognitive radio to scan the spectrum and adaptively switch channels to mitigate secondary user's link degradation. Channel reservation mechanism to reduce the collision when forced termination occurs for the cognitive user's communication. The protocol is employed in decentralized multi channel network with continuous spectrum sensing and scanning capability. The channel status table and reservation information is embedded in the control messages and exchanged with communicating nodes and then announced to the neighbor's nodes. A novel idea is to include the reservation information for the backup channel into the control messages of the current transmission instead of sending an extra control packet, and to use the duration identification (ID) to decide when the currently occupied channel will be released. Thus, the reservation information can be heard by all nodes within the transmission range of the sender with little extra overhead.

Cognitive radio with the spread spectrum modulation techniques, provide a highly secure communication format resistant to deliberate narrowband jamming and other obstruction tactics. Spread spectrum technique, has unique feature to make the data look like noise, and secure in the sense that the jamming and the interfering elements are unable to distinguish mixed data with noise being sent over the channel and prevent from eavesdropping. Different encryption techniques such as integrated key encryption public key encryption (RSA, Elliptic, SHA) and private key encryption (DES, Triple DES, AES) algorithms, and integrated trust computing with public cryptography are useful with cognitive radio to provide a form of secure communication. These encryption algorithms make sure that the key that is used at the transmitter side should be provided by the receiver for correct information retrieval and hence ensures the security and also prevent the malicious users from taking control over the system, blocking the access to other secondary users.

## 2. Related Work

Opportunistic MAC for CR network coexistence with any primary network is a discrete protocol with the advantages of fast-employment and flexibility [1]. The major issue is MAC cannot realize the physical layer status. Furthermore, Multiple channel cognitive radio with backup channel reservation MAC protocol facilitate quick and smoother shift to a reserved channel before the secondary link degrades due to primary user arrival [2] . The protocol significantly improves the network throughput; keep the privilege to the licensee and utilizes the available free channel perfectly with less packet exchange overhead. The simplicity feature makes OC-MAC a more feasible solution than other schemes even without synchronization. But, the major drawback is the physical layer performance degradation due to the lack of co-operation between PHY and MAC layers and lack of security check during the communication [3]. Additionally, OFDM Based Framework for cognitive radio network (OFDM-Based CRN) is deployed in cross-layer design in CRN [4]. The optimization of joint power control and link scheduling is done on the cross-layer architecture of OFDM-based CRN. The dynamic channel assignment within power constraints and the transmission power reduce the interference and improve performance. But, the channel collision is not addressed specifically to optimize the instantly available channel bandwidth. For security, secured and reliable CR communication solutions include the integration of the merits of spread spectrum modulation, using integrated key encryption algorithms as well as trust computing with encryption, and its potential to switch over various frequency bands [5] ,[6].

A fundamental component of the DDCE scheme for complexity reduction in channel estimation for cross layer based CR is a posteriori Least Squares (LS) temporal estimator of the CR-OFDM subcarrier-related Frequency-Domain Channel Transfer Function (FD-CTF) coefficients [7]. The precision of the consequential secular estimates is normally improved using either one or two-dimensional interpolation using both the time and the frequency-domain correlation between the adjacent FD-CTF coefficients. The LS-based secular FD-CTF estimator, with the energy of the transmitted subcarrier related symbols varies as a function of both the modulated sequence and the alternative of the potentially non-constant modulus modulation scheme itself. Minimum Mean Square Error (MMSE) Channel Impulse Response (CIR) estimator-aided DDCE method causes to the complexity reduction in channel estimation which is comparable with LS.

The drawbacks of high SNR performance limitation for MMSE, limitation on the spacing/number of pilots can be mitigated by the finite delay spread of the channel and developing a low complexity algorithm capable of estimating the channel from part of the carriers only [8]. ML estimator can be interpreted as a transformation from frequency domain to time domain and back to frequency. The actual estimation can be done in the time domain, where the number of parameters such as the channel length is small. The estimator can be achieved by minimizing a quadratic criterion, which, combined with the small number of parameters, for lower complexity algorithm with Pilot Symbol Assisted Modulation (PSAM) and Constrained Least Squares (CLS).

## 3. Secured Distributed Cognitive MAC

The implementation of the secured distributed cognitive MAC and complexity reduction in channel estimation is depicted in Fig. 1. When the user application received by the CR manager (CRM) for the link request and processing delay constraint, it performs initial step of authentication using security policy and then forwards the authenticated link request to the Channel Sense and Allocation (CSA) for available free channels. The CSA consists of direct status check of the noise and interference measurement module for probable channel interference power to to detect the free or idle channels. From the response of the noise and interference sensing module located in the PHY layer, the channel sensing and allocation module selects entirely free channel for both authenticated transmitter and receiver by using Channel Endorsement and Transmission Policy (CETP) and sends the detected free channels with their respective identifications to the CRM. This results that the CRM has now the complete information regarding the available free channels for the authenticated requesting applications users. The CRM requests the power aware scheduling module for transmit power limits on each of available free channels. The CRM also sends information about the delay constraints for the requesting application users. The scheduling module recommends to the interference measuring for signal-to-interference and noise ratio needed for joint power control and link scheduling. The scheduling module divides the MAC layer frames into sub-frames depending on delay constraints of authenticated requesting applications and assigns a group of links to each sub-frame. The module also allocates a group of transmit powers based on the delay constraints. The scheduling module sends the information to the sub-frame and probable group of transmit power allocation to each sub-frame. Finally, the CRM checks with the channel estimation for probable error rates and channel fading conditions based on the information received from the power scheduling module. The channel estimation is described in section -4 in details. Then, the CSA allocates the power constrained links to the authenticated application determined by the CRM and update security authentication at some periodic time interval for robustly secured communication.

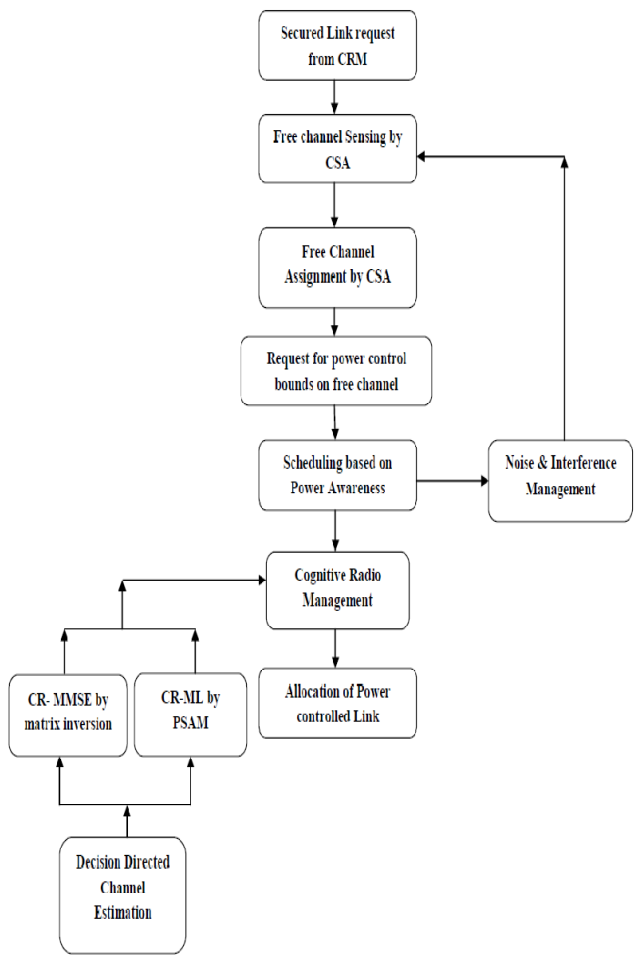

Fig. 1 Distributed Cognitive MAC and Complexity Reduction in Channel Estimation

3.1 Security Policy

The security scheme is including the authentication of CR before giving an exclusive access to spectrum or link and the authentication of users before communication and interval after communication. This addresses the mis-detections and false alarms issues. Security is deployed in three steps to make robustly secured in MAC layer and PHY layer listed as follows:

Step-1: Digital signatures are used in authenticating a CR,

Step-2: Integrated Trust computing and Public Cryptography technique is used in authenticating CRs in communication [6].

Step-3: Ensure security during the interval of communication between users.

In step-3, CR user estimates a maximum probable number of free sub-bands available for its communication using preliminary knowledge. The central idea of computing this probability is to compute the probable number of free sub-bands over which a cognitive radio switches during its communications. A cognitive user uses this information along with the information obtained from the dedicated database to comprehend the sequence of switching pattern for its communication. Once the two cognitive users have authenticated themselves to be genuine, the transmitting user conveys the switching pattern to its intended receiver. This procedure ensures that only the intended users are aware of the switching pattern of frequencies.

3.2 Channel Endorsement and Transmission Policy

For channel endorsement and transmission policy, when a CR node *A* wants to transmit data to CR node *B*, *A* updates its Channel State Table at first. If a free channel *y* becomes busy after update, *A* resets the timer of y which is represented in equation (1) .

$$\text{Timer}(y) = \text{pkt\_size}(y)/\text{rate}(y) \qquad (1)$$

This sensing makes sure that there is not unexpected present of PUs because PUs don't send any message on control channel where the messages are used for recording data channels' status. Generally, a free channel turns to be busy after sensing, when it is occupied by PUs. It is necessary to initiate the available timer as packet transmission duration.

When $B$ received RTS from $A$, it updates the Channel State Table at first. Then $B$ finds the overlapping channel set between $A$'s and $B$'s available lists and uses the evaluating function to score channels. For each channel $x$, the most appropriate channel is the one over which CR pairs can send most packets cumulatively as follows:

$$\text{Max}(\text{Min}_x\{\text{pkt\_num}_{tran}(x), \text{pkt\_num}_{recv}(x)\}) \qquad (2)$$

$$\text{pkt\_num}(x) = [\beta \times \{\text{slot}(x) - LC/\text{pkt\_size}(x)\}] \qquad (3)$$

For the maximum slot number, Equation [2] calculates the maximum cumulatively packet number for distributed MAC. The maximum slot number of channel $x$, $slot(x)$, is given as:

$$1 - (1 - U_t(x))^{slot(x)} < \text{threshold}(x) \qquad (4)$$

$$U_t = \alpha U_{t-1} + (1 - \alpha)\hat{U} \qquad (5)$$

where, $U_{t-1}$ is the exact utilization of last time slot $t-1$ acquiring by physical layer.

$\hat{U}$ is the average experienced utilization of the past.

If there are more than one channel having the same pkt_num, it will select the one with maximum transmission rate. If there still are more than one channel have the same pkt_num with the same rate, randomization is used for decision. After making choice, $B$ sends CTS to $A$ with selected channel index and maximum packet number.

Under the condition that there is no available channel for $B$ or maximal packet number is zero, $B$ do nothing, and $A$'s timer for CTS will expire. After receiving CTS from $B$, $A$ sends CRTS to $B$ for confirmation; at the same time, all CR neighbors of $A$ are also informed of the upcoming transmission over channel $x$ and update their Channel State Tables for channel.

## 4. Complexity Reduction in Channel Estimation

The Decision Directed Channel Estimation (DDCE) scheme is preferred for CR-OFDM rather than purely pilot assisted channel estimation methods because it deploys both information symbols and the pilot symbols for channel estimation. The favorable case for DDCE is in the absence of transmission errors where the full pilot information can be used to detect subcarrier symbols as a posteriori reference signal which drastically reduce the number of pilot symbols required.

Least Square method has low complexity but the performance is degraded due to the energy-fluctuations associated with the near-Gaussian distributed subcarrier-related samples which renders the LS estimator is the worst for decision directed channel in CR-OFDM. On the other hand, MMSE in DDCE has good performance but high complexity over LS which can be mitigated by passing the computationally intensive matrix inversion operation. In addition, the Maximum Likelihood (ML) DDCE for CR-OFDM is optimized for low complexity by novel time-frequency interpretation.

4.1 Complexity Reduction in MMSE SS-CIR Estimator

The block diagram of the channel estimation method consists of a posteriori decision-directed CIR estimator, followed by a priori CIR predictor as shown in Fig. 2. The posteriori SS-CIR estimator inputs are the frequency-domain signal y[n] and the decision-based transmitted symbol estimate x[n]. The transformation from the frequency to time domain is performed within the CIR estimator and its output is a posteriori estimate x[n, k] of the SS-CIR taps, which is fed into the low-rank time-domain SS-CIR tap predictor to produce *a priori* estimate h[n+1, l], l = 0, 1, … ,$K_0$ − 1 of the next SS-CIR on a SS-CIR tap-by-tap basis. Finally, the predicted SS-CIR is converted to the Frequency Domain-CTF at the FFT block. The resultant FD-CTF is employed by the receiver for detecting and decoding the next OFDM symbol. The discrete baseband model of the OFDM system can be given by:

$$y[n, k] = H[n, k]x[n, k] + w[n, k] \quad (6)$$

for k = 0, . . . ,K − 1 and all n, and y[n, k], x[n, k] and w[n, k] are the received symbol, the transmitted symbol and the Gaussian noise sample respectively, corresponding to the $k_{th}$ subcarrier of the $n_{th}$ OFDM symbol. Moreover, H[n, k] is the complex channel transfer function (CTF) coefficient associated with $k_{th}$ subcarrier and time instant n. For an M-QAM modulated OFDM system, x[n, k] corresponds to the M-QAM symbol accommodated by the kth subcarrier. In OFDM systems using a sufficiently long cyclic prefix and adequate synchronisation, the discrete CTF can be formed as follows. For k = 0, 1, …., K − 1, where h[n, l] = h(nT, lT/K) is the sample-spaced CIR having significant nonzero tap values only at sample-spaced raster positions and $W_K$ = exp(−j2π/K).

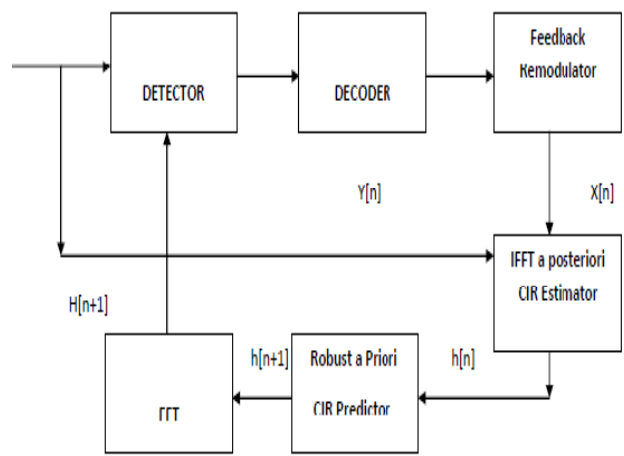

Fig. 2 Channel estimator constituted by an a posteriori decision-directed CIR with priori CIR

$$H[n,k] = H(nT, k\Delta f) = \sum_{l=0}^{k_0-1} W_K^{kl} h[n,l], \quad (7)$$

By substituting $\hat{h} = \sigma_w^2 I + diag(\sigma_i^2) \hat{W}^H diag$ from Frequency Domain CTF Equation (1) into (2);

$$y[n,k] = \sum_{l=0}^{K0-1} W_K^{kl} h[n,l] x[n,k] + w[n,k]. \quad (8)$$

The SS-CIR taps h[l] are assumed to be uncorrelated complex-Gaussian distributed variables having a zero mean and a covariance matrix $C_h$ = diag ($\sigma^2$). The MMSE estimator of the SS-CIR taps h[n, l] of the linear vector model described by equation (8) is given as follows:

$$\hat{h} = \left(\sigma_w^2 I + diag(\sigma_i^2) \hat{W}^H diag(|\hat{x}[k]|^2) W\right)^{-1} \times diag(\sigma_i^2) \hat{W}^H diag(\hat{x}^*[k]) y \quad (9)$$

The direct MMSE for estimation of the SS-CIR taps h[n, l] involves an inversion of a time-variant matrix which imposes a relatively high computational complexity as compared to the least square. In order to reduce the allied computational complexity, a two-step reduced complexity SS-CIR estimator invoking a method, which bypasses the computationally intensive matrix inversion operation. The channel Transfer function of MMSE can be given as follows:

$$H_{MMSE}^n[n,k] = \frac{\hat{x}^*[n,k] \times y[n,k]}{|x[n,k]|^2 + \frac{\sigma^2}{\sigma_H^2}} \quad (10)$$

MMSE CTF from equation (10) may be modelled as complex Gaussian-distributed variables having a mean identical to that of H[n, k], which represents the actual FD-CTF coefficients encountered and a variance of $\sigma_2^v = \sigma_2^{H_{MMSE}}$.

$$H_{MMSE}^n[n,k] = H[n,k] + v[n,k], \quad (11)$$

where, v[n, k] represents the i.i.d. complex-Gaussian noise samples having a zero mean and a variance, and substituting equation (7) into (11) yields as follows:

$$H_{MMSE}^n[n,k] = \sum_{l=0}^{K0-1} W_K^{kl} h[n,k] + v[n,k], \quad (12)$$

Furthermore, the MMSE estimator of the SS-CIR taps h[n, l] of the linear vector model described by equation (12) is given as follows:

$$\hat{h} = \hat{W}^H \frac{H_{MMSE}^n}{\left[C_v\left(\frac{1}{C_h}\right) \hat{W}^H\left(\frac{1}{C_v}\right) W\right]} \quad (13)$$

where, we define $C_h$ and $C_v$ as the covariance matrices of the SS-CIR vector h and the scalar-MMSE FD-CTF estimator's noise vector v, respectively. The elements of the noise vector v are assumed to be complex-Gaussian i.i.d. samples and therefore we have $C_v = \sigma_v^2 I$. On the other hand, as follows from the assumption of having uncorrelated SS-CIR taps, the SS-CIR taps' covariance matrix is a diagonal matrix $C_h$ =diag ($\sigma^2$) where $\sigma^2_l$ = E {h[n, l]2}. Substituting $C_h$ and $C_v$ into Equation (8) yields;

$$\hat{h} = \left(diag\left(\frac{1}{\sigma_i^2}\right) + \left(\frac{1}{\sigma_v^2}\right) \hat{W}^H W\right)^{-1} \hat{W}^H \left(\frac{1}{\sigma_v^2}\right) H_{MMSE}^n$$

$$\hat{h} = diag\left(\frac{\sigma_i^2}{\sigma_v^2 + K\sigma_i^2}\right) \hat{W}^H H_{MMSE}^n \quad (14)$$

Finally, upon substituting Equation (10) into Equation (12), it gives a scalar expression for the Reduced-Complexity (RC) a posteriori MMSE SS-CIR estimator in the form of;

$$\hat{h}[n,l] = \left(\frac{\sigma_l^2}{\sigma_v^2 + K\sigma_l^2}\right)\left(\sum_{k=0}^{i-1} W_k^R \frac{\overset{\times}{x}[n,k] \times y[n,k]}{|x[n,k]|\left(2 - \frac{\sigma_v^2}{\sigma_H^2}\right)}\right) \quad (15)$$

The Complexity Reduced MMSE (CR-MMSE) estimator of equation (15) does not require any matrix inversion and therefore it has outstanding performance as the allied complexity is equivalent to the general LS estimator.

4.2 Complexity Reduction in Maximum Likelihood

The complexity of ML solution can be applied to a combination of pilot symbol assisted modulation (PSAM) and decision-directed channel estimation (DDCE) OFDM system as Complexity Reduced Maximum Likelihood (CR-ML) The pilot symbols along with decisions taken on the other carriers, then r = h + v remains valid, with a given $C_{vv}$, which leads to H =$P_{Fh}$ y. The combination of PSAM and decision-feedback is desirable as the increment in complexity can be dealt. The complexity of the ML estimator can be significantly minimized both for spectral shaping and PSAM systems. This low complexity depends on the time-frequency interpretation and pruning of the FFT/IFFTs.

Regarding spectral shaping, manipulating $P_{Fuh}$ as a low rank matrix of rank $N\check{}_h$ considering its hermiticity, $P_{Fuh} = VV^H$ where V is a matrix of size $N_c$ x $N\check{}_h$ and can be computed easily. The complexity of computing the estimator is 2 $N_u$ x $N\check{}_h$ complex multiplications for the general ML estimator. This complexity can be reduced by using the time-frequency interpretation. The projection operation can be expressed by the cascade of two partial FFTs, weighted by a $N\check{}_h$ x $N\check{}_h$ matrix assuming that all carriers are used as pilots, called an identity matrix.

When using Pilot Symbol Assisted Modulation, a comb spectrum has to be measured, and only the teeth of the comb are used for the FFTs. The DFT can be computed with $N_c/4 + N\check{}_h/2 + \log_2 N\check{}_h \cdot N\check{}_h$ complex multiplications, which refers a large gain for a large number of carriers [8]. The complexity for FFT-based solutions is much lower than for the singular value decomposition (SVD) based approach, both for spectral shaping and PSAM systems. Futhermore, the ML algorithm can efficiently work with a significantly smaller $N\check{}_h$ than the LMMSE and LS, which yields to comparatively larger gain.

The major inspiration to the complexity is due to the weighting matrix for a relatively large number of pilot carriers. Rigorously, pure PSAM with regularly spaced pilot carriers can easily be depicted that the weighting matrix $(F_{uk}^H F_{uk})^{-1}$ is proportional to the identity matrix, and complexity is even lower. This is the special case of frequency correlation interpretation.

## 5. Simulation and Performance Evaluation

The Simulation is conducted setting the total licensed frequency 600 MHz and 100 allocated channels at the rate of 6MHz. The maximum licensed primary users are 100 and spectrum hole is generated randomly for secondary users in the absence of incumbents. Fig. 3 represents WRAN spectrum considering video signal spectrum for instant 58 users including both incumbents as well as secured CR users in ad hoc manner without distributed MAC. Fig. 4 illustrates 58 active primary users SNR, whereas Fig. 5 refers the dropped or unsecured 22 secondary users and Fig. 6 refers authenticated 20 secondary user SNR at corresponding base stations. The power controlled link without using Transmission policy and security policy is represented in Fig. 7 and found adverse power assigned in links non-uniformly which degrades the network performance.

Furthermore, the Fig. 8 Fig. 9 and Fig. 10 show that the transmitted packets, received packets and their corresponding packet ratio at different channel sub-bands deploying Channel Endorsement and Transmission Policy. The channels are selected based on the joint information of the utilization function and slots to determine the cumulative packet distribution in distributed MAC and compared to threshold. The best channels are having maximum good put or higher packet transmission, in which even the minimal data transmitted without loss. The Power Controlled Link is assigned to best channel application by CSA after channel estimation in Fig.11 and their corresponding optimized throughput up to 75% in Fig. 12. This concurs that the CETP and SP with cross-layer based Power Controlled Links (PCL) provide the optimal number of maximum transmission oriented channels and provide higher throughput as compared to without PCL.

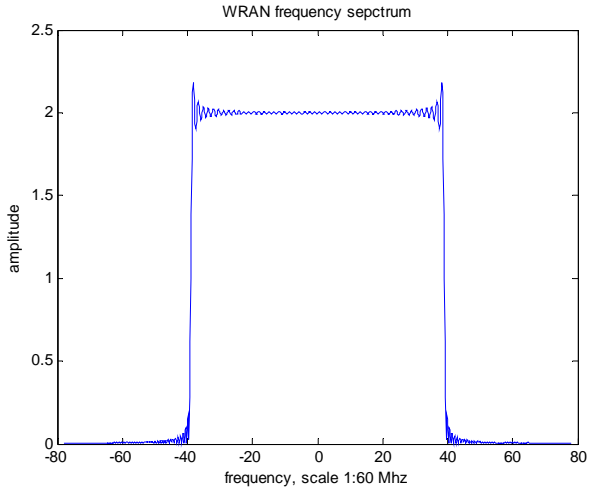

Fig. 3 Ad hoc WRAN Spectrum 600 MHz (scale 1:60)

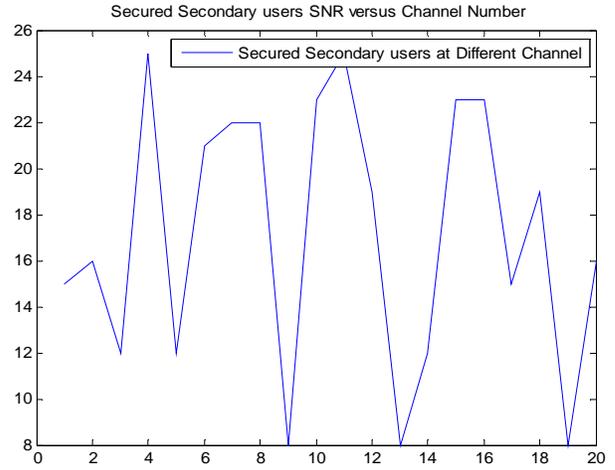

Fig. 6 SNR in authenticated Secondary user channels

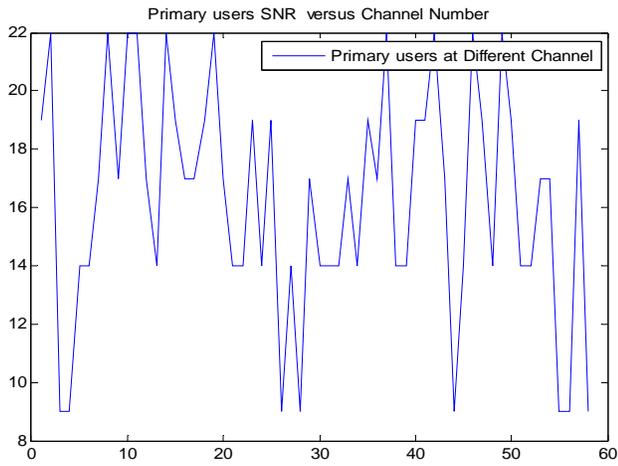

Fig. 4 SNR in Primary user channels

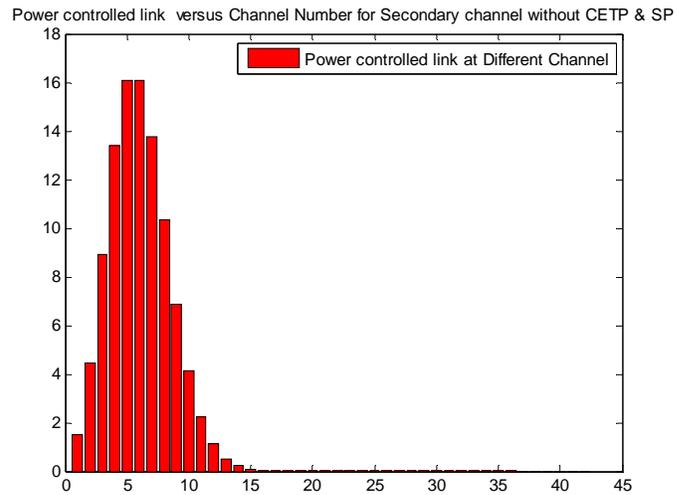

Fig. 7 Power controlled links without Transmission and Security policy

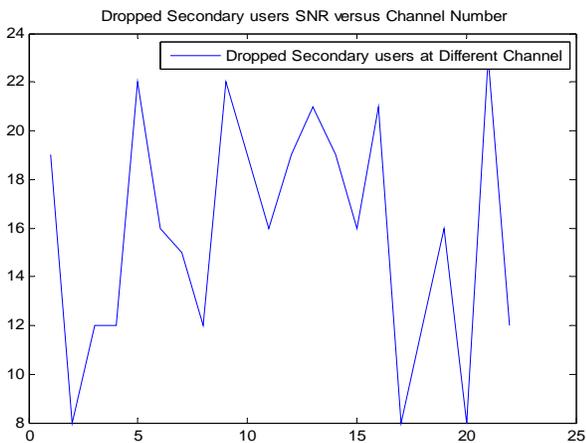

Fig. 5 SNR in Dropped (unsecured) Secondary user channels

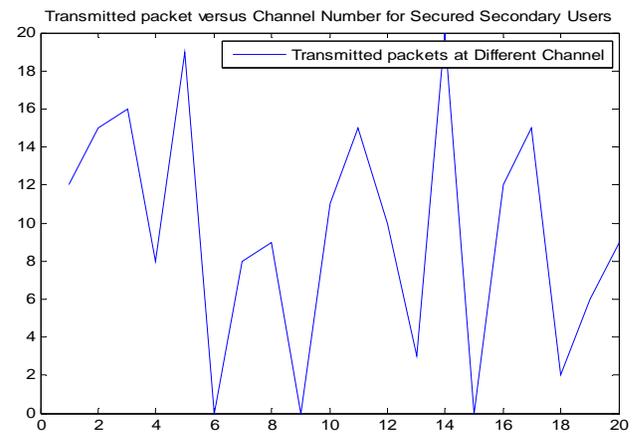

Fig. 8 Transmitted packets in Secondary channels

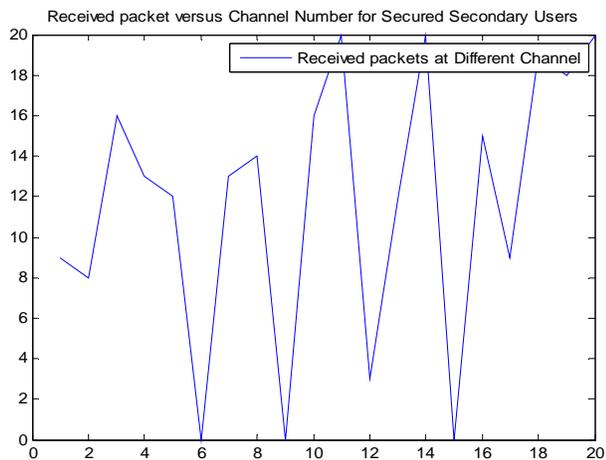

Fig. 9 Received packets in Secondary channels

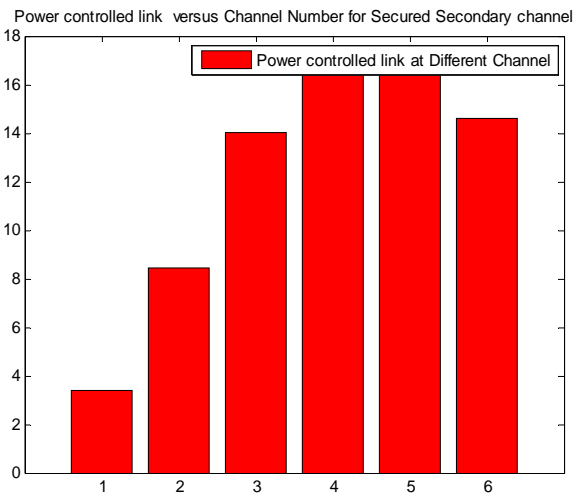

Fig. 11 Power Controlled Secondary Channels deploying UTEP and SP

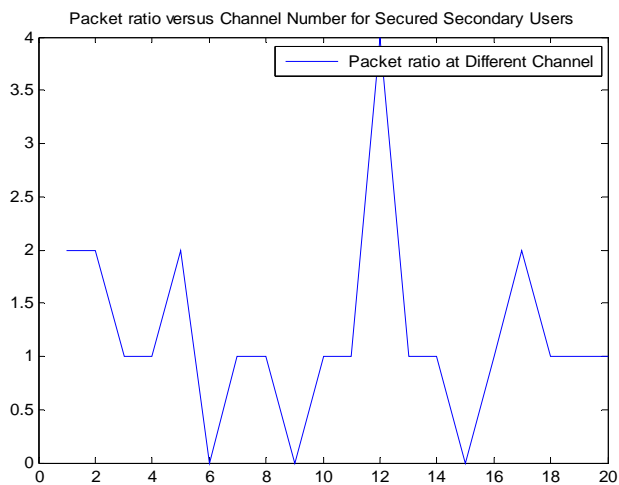

Fig. 10 Packets Ratio of Transmitted and Received packets in Secondary channels using CET

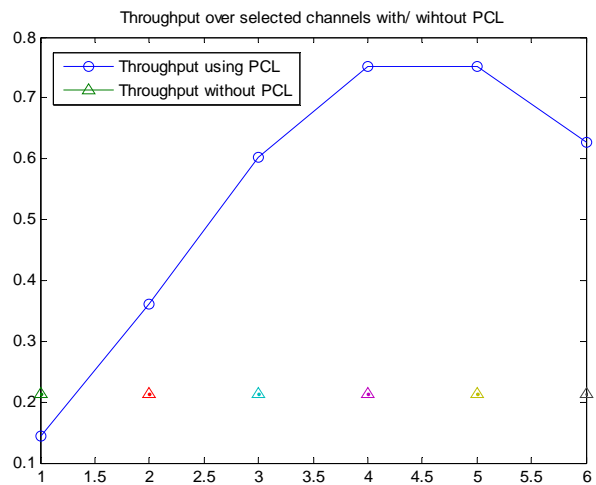

Fig. 12 Throughput over selected channels with/without power Constrained links

Regarding complexity reduction in channel estimation, CR based on OFDM with DDCE specifications including the various parameters are as given below in Table-1. From simulation, SER is optimized at $10^{-1.5}$ for CR-MMSE, $10^{-1.45}$ for MMSE and $10^{-1.35}$ for LS as shown in Fig. 13. In other words, SER is minimized in CR-MMSE and outperforms over MMSE and LS. In conclusion, MMSE has comparatively higher complexity over Least Square but the performance is better. The LS estimator is the worst option for decision directed channel in OFDM. Complexity reduced MMSE in DDCE has better performance than MMSE and low complexity as LS which is achieved by passing the computationally intensive matrix inversion operation.

The behavior of the ML estimator at low SNR gives some insight on the influence of $N\hat{}h$. As moving from higher to lower $N\hat{}h$ tends to better performance. The values of the channel impulse response beyond the lower $N\hat{}h$ are below the noise level, so that their estimation introduces more noise than relevant information about the channel. As for the spectral shaping system, the flooring effect of the LMMSE estimator is essentially limiting its effectiveness at high SNR's. Simulation results demonstrate that the LMMSE suffers from a threshold effect at high SNR and gives SER converged to $10^{-1.4}$ whereas improving performance $10^{-1.47}$ and $10^{-1.49}$ in Complexity Reduced ML (CR-ML) and ML as shown in Fig. 14. $N\hat{}h$ must be 2 to 4 times larger for LMMSE to trade off SER and SNR than for ML.

Table- 1 Simulation Parameters for Channel Estimation

| Simulation Parameters | Value |
|---|---|
| Modulation | BPSK |
| Number of used subcarriers (nDSC) | 64 |
| Channel Model | AWGN |
| Channel B/W | 1000 kHz |
| CIR taps | 4 |
| Max Delay Spread | 30 μs |
| Total Symbol duration (Ts) | 120 μs |
| Twiddle Matrix size | 64 x 64 |

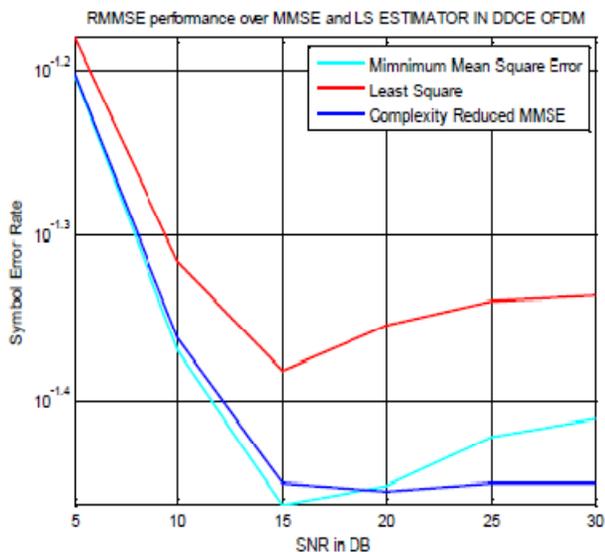

Fig. 13 SNR versus SER plot for MMSE, LS, CR-MMSE

## 6. Conclusion

Secured Distributed MAC and complexity reduction in channel estimation for Cross-layer based Cognitive radios provide outstanding performance in terms of robustness, SER, joint power control and link scheduling with optimum opportunistic MAC. This may slightly increase the processing overheads and resource consumption as compared to without cross layer approach but it has significant achievement in terms of throughput, quality of service, interference mitigation and better information rate. On the other hand, CR-MMSE has features of complexity reduction by allowing to pass through the computationally intensive matrix inversion operation which demonstrates outstanding SER performance over MMSE and LS in sample spaced channel impulse response (SS-CIR) of DDCE OFDM. Furthermore, CR-ML featured by partial IFFT/FFT with Pilot Symbol Assisted Modulation (PSAM) provides the significant SER performance over MMSE and ML in Cross-layer based Cognitive radio Networks.

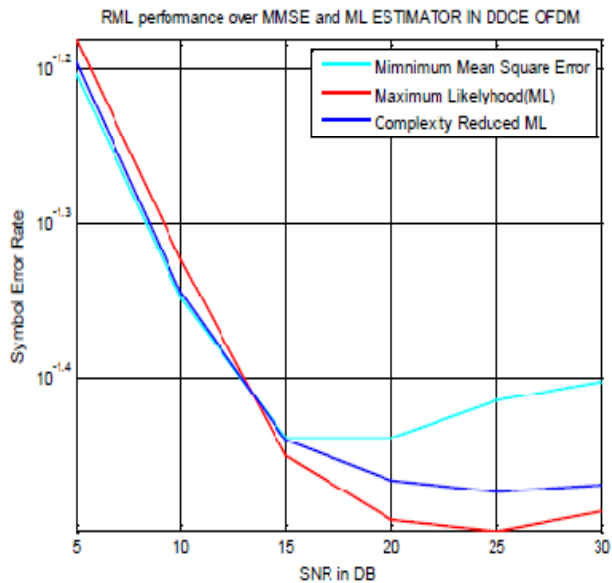

Fig. 14 SNR versus SER plot for MMSE, ML, CR-ML

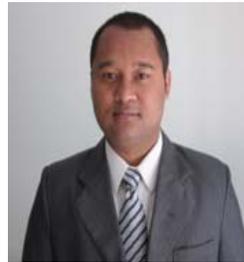

**Dr. Niraj Shakhakarmi** worked as a Doctoral researcher from 2009 to 2011 in the US ARO (Army Research Office) funded Center for Battlefield Communications (CeBCom) Research, Department of Electrical and Computer Engineering, Prairie View A&M University (Texas A&M University System). He received his B.E. degree in Computer Engineering in 2005 and M.Sc. in Information and Communications Engineering in 2007. He has accomplished Ph.D in Electrical & Computer Engineering in 2011 from Prairie View A&M University, Houston, USA. His research interests are in the areas of Cognitive and Cooperative Radio Networks, Wavelets applications, Digital Image Processing, Secured Position Location & Tracking (PL&T), Mobile Ad hoc Networks, 4G Networks, Satellite Networks, QoS & Network Security, and Color Technology. He has published WSEAS journal, IJCSI journal, ICSST conference, Elsevier conference, and several other papers are under review in IEEE and WSEAS journals and conference papers.